\documentclass[12pt, prd, showpacs]{revtex4}
\usepackage{amssymb}
\usepackage{amsmath}

\input{tcilatex}

\begin{document}

\title{Schwarzschild black hole as accelerator of accelerated particles}
\author{O. B. Zaslavskii}
\affiliation{Department of Physics and Technology, Kharkov V.N. Karazin National
University, 4 Svoboda Square, Kharkov 61022, Ukraine}
\affiliation{Institute of Mathematics and Mechanics, Kazan Federal University, 18
Kremlyovskaya St., Kazan 420008, Russia}
\email{zaslav@ukr.net }

\begin{abstract}
We consider collision of two particles near the horizon of a nonextremal
static black hole. At least one of them is accelerated. We show that the
energy $E_{c.m.}$ in the center of mass can become unbounded in spite of the
fact that a black hole is neither rotating nor electrically charged. In
particular, this happens even for the Schwarzschild black hole. The key
ingredient that makes it possible is the presence of positive acceleration
(repulsion). Then, if one of particles is fine-tuned properly, the effect
takes place. This acceleration can be caused by an external force in the
case of particles or some engine in the case of a macroscopic body
("rocket"). If the force is attractive, $E_{c.m.}$ is bounded but, instead,
the analogue of the Penrose effect is possible.
\end{abstract}

\maketitle

\section{Introduction}

During last decade, much attention was focused on high energy collisions
near black holes. It was stimulated by the observation that, under certain
conditions, collisions of free falling particles near the horizon can lead
to the unbounded energy $E_{c.m.}$ in the center of mass frame. This is the
Ba\~{n}ados-Silk-West (BSW) effect \cite{ban} that occurs due to the
existence of special fine-tuned (so-called critical) trajectories. For
rotating black holes the corresponding relation between the energy at
infinity $E$ and angular momentum $L$ reads $E=\omega _{H}L$, where $\omega $
is the metric coefficient responsible for rotation and proportional to the
angular momentum of the black hole, subscript "H" refers to the horizon. The
counterpart of the BSW effect can happen also near static but electrically
charged black holes \cite{jl}, with the difference that instead of
geodesics, particles move under the action of the electrostatic field. Then,
the critical trajectories are characterized by equality $E=\frac{qQ}{r_{+}}$%
, where $q$ and $Q$ are the charges of a particle and black hole,
respectively, $r_{+}$ being the horizon radius. In other words, for the BSW
effect to occur, a black hole should be rotating or/and charged. If one puts 
$\omega =0$, $Q=0$, this kills the effect in both cases. Correspondingly,
the Schwarzschild black hole was unable to serve as particle accelerator to
unbounded $E_{c.m.}$ (We put aside the case when colliding particles are
spinning.). More precisely, if two particles of equal mass $m$ collide in
the Schwarzschild background, $E_{c.m.}$ $\leq 2\sqrt{5}m$ \cite{baush}.

It turns out, however, that modification of the set-up changes the situation
drastically, if instead of geodesic motion, we consider motion under the
action of some finite force. In contrast to motion of a particle in the
Reissner-Nordstr\"{o}m space-time, this force can have external source. We
will see that for the motion under such a force the analogue of the BSW
effect is possible even in the Schwarzschild background, i.e. for a
nonrotating and electrically neutral black hole.

The role of a finite force was discussed earlier \cite{rad}, \cite{ne} but
just for rotating and/or charged black holes. It was implied in the
aforementioned papers that the effect under discussion existed without this
force. The question was whether or not the force spoils the effect. It was
found that (under some additional weak restrictions on the behavior of the
force) the BSW\ effect survives. Meanwhile, now the existence of such a
force is not a potential obstacle but, by contrary, it is the only cause of
the effect.

In what follows, we use the geometric system of units in which fundamental
constants $G=c=1$.

\section{Basic formulas}

Let us consider the black hole metric%
\begin{equation}
ds^{2}=-fdt^{2}+f^{-1}dt^{2}+r^{2}(d\theta ^{2}+\sin ^{2}\theta d\phi ^{2})%
\text{,}  \label{met}
\end{equation}%
where the horizon is located at $r=r_{+}$, so $f(r_{+})=0$. We will consider
pure radial motion with the four-velocity $u^{\mu }$ and four-acceleration $%
a^{\mu }$ with 
\begin{equation}
a_{\mu }a^{\mu }\equiv a^{2},  \label{a2}
\end{equation}%
where by definition $a\geq 0$. Then, it is convenient to use the expression
for the acceleration in the simple form (see e.g. eqs. 16, 17 in \cite{sym}
and \cite{pk}): 
\begin{equation}
a^{t}=f^{-1}u^{r}\frac{d}{dr}(fu^{t})\text{,}  \label{at}
\end{equation}%
\begin{equation}
a^{r}=fu^{t}\frac{d}{dr}(fu^{t})\text{.}  \label{ar}
\end{equation}%
In \cite{sym}, \cite{pk}, eqs. (\ref{at}), (\ref{ar}) were exploited for
investigation of trajectories with the constant acceleration in the
Schwarzschild space-time but they are valid in a more general case as well.

By substitution into (\ref{a2}), one finds for a particle having the mass $m$%
:%
\begin{equation}
mu^{t}=\frac{X}{f}\text{,}  \label{ut}
\end{equation}%
\begin{equation}
mu^{r}=\sigma P\text{, }P=\sqrt{X^{2}-m^{2}f}\text{,}  \label{ur}
\end{equation}%
$\sigma =\pm 1$.%
\begin{equation}
X=m\beta (r)+E\text{.}  \label{X}
\end{equation}%
Here, $E$ is a constant of integration, 
\begin{equation}
\beta =\delta \int^{r}dr^{\prime }a(r^{\prime })\text{,}  \label{beta}
\end{equation}%
$\delta =\pm 1$. The forward-in-time condition $u^{t}>0$ requires 
\begin{equation}
X\geq 0.  \label{forw}
\end{equation}%
It follows from (\ref{ar}) that 
\begin{equation}
a^{r}=\frac{aX}{m}\delta \text{,}
\end{equation}%
so $\delta =signa^{r}$.

If $a(r)$ tends to zero at infinity rapidly enough, it is convenient to
choose the limit of integration in such a way that%
\begin{equation}
\beta =-\delta \int_{r}^{\infty }dr^{\prime }a(r^{\prime })\text{,}
\end{equation}%
so $sign\beta =-\delta $. If $a=const$ we can choose $\beta =\delta ar$ and
we return to the trajectory considered in \cite{pk}. Then, $sign\beta
=+\delta $. If, additionally, $f=1$, we obtain motion along the Rindler
trajectory with $r=\frac{\cosh a\tau -\frac{E}{m}}{a}$, $t=\frac{\sinh a\tau 
}{a},$ where we put $\delta =+1$ to have $r>0$ for large $\tau $. For the
Reissner-Nordstr\"{o}m (RN) case $a=\frac{\left\vert qQ\right\vert }{mr^{2}}$%
, where $q$ is the particle charge, $Q$ is that of a black hole. The BSW
effect exists in the RN background if for trajectories with $E>0$ we take $%
\delta =+1$, $\beta <0$, $qQ>0$ \cite{jl}. Then, 
\begin{equation}
a^{r}>0,  \label{pos}
\end{equation}%
so we have repulsion between a particle and a black hole.

Let particles 1 and 2 move from infinity and collide in some point $r_{0}$.
The energy in the center of mass frame%
\begin{equation}
E_{c.m.}^{2}=-(m_{1}u_{1}^{\mu }+m_{2}u_{2}^{\mu })(m_{1}u_{1\mu
}+m_{2}u_{2\mu })=m_{1}^{2}+m_{2}^{2}+2m_{1}m_{2}\gamma \text{,}
\end{equation}%
where $\gamma =-u_{1\mu }u^{2\mu }$ is the Lorentz factor of relative
motion. It follows from the above equations that%
\begin{equation}
\gamma =\frac{X_{1}X_{2}-\sigma _{1}\sigma _{2}P_{1}P_{2}}{m_{1}m_{2}f}.
\label{lor}
\end{equation}

One reservation is in order. We assume that the background is described by
the spherically symmetric metric (\ref{met}). In doing so, backreaciton of a
particle and external sources on the metric is neglected. For the case of
the Schwarzschild metric this implies that $m\ll M$, where $M$ is a black
hole mass. Also, we assume that an acceleration is small enough, $a\ll
M^{-1}.$ The full self-consistent picture that takes into account
backreaction on the metric is quite complicated even without an external
force (see, say. Ch. 4 in \cite{fn} and references therein) and is beyond
the scope of our paper.

\section{High energy collisions}

Now, the standard classification applies. A particle is called usual if $%
X_{H}>0$ if it is separated from zero and it is critical if $X_{H}=0$. This
is possible if $\delta =+1$, so $\beta <0$ for a particle coming from
infinity, when $E>0$. Then, for the critical particle%
\begin{equation}
E=\left\vert \beta (r_{+})\right\vert \text{.}  \label{cr}
\end{equation}%
Assuming at infinity $f=1$, we require $E>0$ where it has the meaning of the
Killing energy. We also assume that both particles move towards the horizon, 
$\sigma _{1}=\sigma _{2}=-1$.

Near $r_{+}$, one has $X\approx X_{H}$ for a usual particle. Near the
horizon, using the main term in the Taylor expansion, we have for the
critical particle, 
\begin{equation}
X(r)\approx b(r-r_{+})\text{, }b>0.
\end{equation}

Then, for collisions of the critical particle and a usual one we obtain that
near the horizon the expression inside the square root in (\ref{ur}) becomes
negative. This is manifestation of the known fact that the critical particle
cannot reach the horizon in the nonextremal case. At first, this property
was considered as an obstacle again the BSW effect \cite{berti}, \cite{ted}.
However, the situation changes if instead of exactly critical particle, we
consider a near-critical one with small but nonzero $X_{H}$. (For the first
time, this idea was realized for the Kerr black hole in \cite{gp}). For such
a particle 
\begin{equation}
X\approx X_{H}+b(r-r_{+})\text{.}  \label{x}
\end{equation}%
Let $X_{H}(r_{0})\approx m_{1}d\sqrt{r_{0}-r_{+}}$, where $d$ is come
constant.

Then, it is the first term in (\ref{x}) which dominates$.$ For the metric
function,%
\begin{equation}
f(r_{0})\approx 2\kappa (r_{0}-r_{+})\text{,}
\end{equation}%
where $\kappa $ is the surface gravity. We assume that near-critical
particle 1 collides with a usual particle 2. As a result, we find from (\ref%
{lor}) that

\begin{equation}
\gamma \approx \frac{D}{\sqrt{(r_{0}-r_{+})}}\text{, }D=\frac{\left(
X_{2}\right) _{H}(d-\sqrt{d^{2}-2\kappa })}{2\kappa m_{2}}\text{,}
\label{ga}
\end{equation}%
where we also assumed that $d>\sqrt{2\kappa }$. Then, taking $r_{0}$ as
close to $r_{+}$ as one likes, we obtain the unbounded growth of $\gamma $
and $E_{c.m.}^{2}$ Eq. (\ref{ga}) can be thought of as a counterpart of eq.
(19) from \cite{gp} where the Kerr metric was considered. Thus there is a
close analogy between our case and the BSW effect near nonextremal black
holes. In particular, now the same difficulties persist that forbid arrival
of the near-extremal particle from infinity because of the potential barrier
typical of any nonextremal black hole (for the Kerr metric, see Figure and
accompanying discussion in \cite{gp42}). Therefore, either such a particle
is supposed to be created already in the vicinity of the horizon from the
very beginning or one is led to exploiting scenarios of multiple scattering 
\cite{gp}.

\section{BSW effect, kinematics and Penrose process}

As is known, kinematically the BSW effect can be explained as collision of a
slow particle with a rapid one having the velocity close to the speed of
light \cite{k}. This explanation applies here as well. A usual particle
approaches the horizon almost with the speed of light (see, e.g. eq. 102.7
in \cite{LL}). However, the presence of nonzero acceleration does not
guarantee by itself that the velocity $V$ of the second particle slows down
to some $V<1$. What is required is the criticality condition (\ref{cr}),
provided $\delta =+1$, $\beta <0$, $a^{r}>0$.

Let, instead, $\delta =-1$, $\beta >0$, $a^{r}<0$, so the combination $%
E+m\left\vert \beta \right\vert $ appears in $X$. Then, (\ref{cr}) is
violated and the BSW effect is absent. However, then another interesting
feature appears. Namely, condition (\ref{forw}) can be satisfied if $E<0$, $%
\left\vert E\right\vert <m\left\vert \beta \right\vert $. (Obviously, a
corresponding particle cannot come from infinity.) In turn, the presence of
negative energy states can lead to the Penrose effect with energy
extraction. This is the counterpart of the pure electric kind of this effect
when rotation is absent \cite{ruf}. Thus there is some complementarity
between the BSW and Penrose effects in our case: either we have high energy
collision but without energy extraction or the possibility of such
extraction but with finite $E_{c.m.}$

The existence of the negative energy states in the case $\delta =-1$ allows
one to intoduce a notion of the generalized ergosphere by analogy with what
occurs for the RN metric \cite{ruf}. Its boundary is defined by equation 
\begin{equation}
\beta =\sqrt{f}\text{.}  \label{bf}
\end{equation}

It is obtained by putting $E=0$ in eqs. (\ref{ur}), (\ref{X}). For the
particular case of the RN metric, $\beta =\frac{\left\vert qQ\right\vert }{mr%
}$ and we return to eq. (12) of \cite{ruf}.

In the present article, we mainly concentrate on the effect of unbounded $%
E_{c.m.}$. Whether or not one can achieve unbounded $E$ as well, is another
interesting question. Here, we restrict ourselves by some general remarks.
There exists a special case of the Penrose process in which the energy of an
outgoing particle is indeed formally unbounded (in the test particle
approximation). It is called super-Penrose process (SPP). For rotating back
holes, the SPP\ is forbidden for neutral particles (see \cite{fraq} and
references therein). However, it is possible in principle for charged
particles in the RN metric \cite{rn} even including its flat limit \cite%
{flat}. For these reasons, one can expect the SPP now as well. However, eqs.
(\ref{X}), (\ref{beta}) with finite $X$ show that this would require
formally unbounded acceleration $a$. As long as we are restricted by $%
a\lesssim M^{-1}$ in the test particle approximation, the expected effect is
big but yet bounded. More careful treatment of this issue with discussion of
possible scenarios requires separate work.

One may also ask, whether it is possible to obtain an effect of high energy $%
E$ by replacing a force exerted on a particle with a rocket-like motion in
the Tsiolkovsky spirit, when acceleration arises due to backreaction of
fuel. The answer seems to be negative since the initial and final energies
including those of thrown parts should coincide but in the Schwarzschild
background there are no negative energies in the absence of an external
force, so there is no final surplus of energy. 

\section{Discussion and conclusions}

What is especially interesting is that the effect under discussion is valid
for the Schwarzschild black hole. It is also worth mentioning that for the
extremal case, both terms inside the square root in (\ref{ur}) have the same
order and the result is similar to that for the RN black hole \cite{jl}.
However, now the acceleration may be caused not by interaction between the
particle charge and that of a black hole but by some external force or the
engine on a rocket in the macroscopic case. The results are valid both for $%
a=const$ and for $a\rightarrow 0$ at infinity.

It is worth mentioning that there is one more but a quite different
situation when high energy collisions also occur in the Schwarzschild
background. This happens if a black hole is immersed in a magnetic field. In
doing so, a particle orbiting around a black hole collides with another one
coming from infinity. The magnetic field is supposed to be rather big to
ensure the innermost stable circular orbit to be located near the horizon 
\cite{fr}. Meanwhile, the mechanism considered by us works for radial motion
and any \textit{finite }$a$ and has quite universal character.

Usually, the factor connected with additional forces (like gravitational
radiation) are referred to as obstacles to gaining large $E_{c.m.}$\cite%
{berti}. To the extent that such influence can be modeled by some force,
backreaction does not spoil the effect \cite{rad}, \cite{ne}. Meanwhile, as
we saw now, in our context the presence of the force not only is compatible
with the BSW effect but it can be its origin!

If a black hole is surrounded by external electromagnetic fields, we can
suppose that the described mechanism promotes high energy collisions near
black holes. The Schwarzschild metric and radial motion give us the simplest
exactly solvable example but it is quite probable that qualitatively the
similar results hold in a more realistic situation as well.

\begin{acknowledgments}
I thank Yuri Pavlov for stimulating comments. This work is performed
according to the Russian Government Program of Competitive Growth of Kazan
Federal University.
\end{acknowledgments}

\end{document}